\documentclass[conference]{IEEEtran}
\IEEEoverridecommandlockouts
\usepackage{cite}
\usepackage{amsmath,amssymb,amsfonts}
\usepackage{algorithmic}
\usepackage{graphicx}
\usepackage{textcomp}

\usepackage{tablefootnote}
\usepackage{multirow}
\usepackage{multicol}
\usepackage{soul,xcolor}
\usepackage[normalem]{ulem}

\graphicspath{{figures/}} 

\begin{document}

\title{Reinforcement Learning-Based Deadline and Battery-Aware Offloading in Smart Farm IoT-UAV Networks}
\author{\IEEEauthorblockN{Anne Catherine Nguyen$\dag$, Turgay Pamuklu$\dag$, \IEEEmembership{Member, IEEE}, Aisha Syed$\ddag$, \\ W. Sean Kennedy$\ddag$, Melike Erol-Kantarci$\dag$, \IEEEmembership{Senior Member, IEEE}}

\IEEEauthorblockA{$\dag$\textit{School of Electrical Engineering and Computer Science,}
\textit{University of Ottawa}, Ottawa, Canada}

\IEEEauthorblockA{$\ddag$\textit{Nokia Bell Labs}\\
Emails:\{anguy087, turgay.pamuklu, melike.erolkantarci\}@uottawa.ca, 
\{aisha.syed, william.kennedy\}@nokia-bell-labs.com}
}
\maketitle
\makeatletter
\def\ps@IEEEtitlepagestyle{%
  \def\@oddfoot{\mycopyrightnotice}%
  \def\@oddhead{\hbox{}\@IEEEheaderstyle\leftmark\hfil\thepage}\relax
  \def\@evenhead{\@IEEEheaderstyle\thepage\hfil\leftmark\hbox{}}\relax
  \def\@evenfoot{}%
}
\def\mycopyrightnotice{%
  \begin{minipage}{\textwidth}
  \centering \scriptsize
Accepted Paper. IEEE policy provides that authors are free to follow funder public access mandates to post accepted articles in repositories. When posting in a repository, the IEEE embargo period is 24 months. However, IEEE recognizes that posting requirements and embargo periods vary by funder. IEEE authors may comply with requirements to deposit their accepted manuscripts in a repository per funder requirements where the embargo is less than 24 months.
  \end{minipage}
}
\makeatother

\begin{abstract}
Unmanned aerial vehicles (UAVs) with mounted base stations are a promising technology for monitoring smart farms. They can provide communication and computation services to extensive agricultural regions. With the assistance of a Multi-Access Edge Computing infrastructure, an aerial base station (ABS) network can provide an energy-efficient solution for smart farms that need to process deadline critical tasks fed by IoT devices deployed on the field. In this paper, we introduce a multi-objective maximization problem and a Q-Learning based method which aim to process these tasks before their deadline while considering the UAVs' hover time. We also present three heuristic baselines to evaluate the performance of our approaches. In addition, we introduce an integer linear programming (ILP) model to define the upper bound of our objective function. The results show that Q-Learning outperforms the baselines in terms of remaining energy levels and percentage of delay violations.
\end{abstract}
\section{Introduction}
\par One of the challenges of agriculture is battling against nature, this includes: pest control, fire control, and monitoring crop growth. A farm consists of hundreds of acres of land and it is difficult to monitor all the crops for pests. Fires can be less detrimental to the crops if they are detected and pinpointed early. With the introduction of smart agriculture, farmers can use internet of things (IoT) devices with cameras to monitor their crops and use image recognition to detect pests, fires, and the growth stages of crops. 

Image classification has emerged as a useful tool for agriculture. Aldabbagh et al. introduced a Deep Learning algorithm to predict the growth stage of chili plants from images in \cite{Aldabbagh2020}. In \cite{Lina2020}, Yu et al. proposed a pest monitoring system that comprised of unmanned aerial vehicles (UAV) with a mounted camera and ground sensors to monitor fields of crops. The UAVs flew across the fields and captured images and performed image classification to detect pest infestations on crops. 

IoT devices are limited in computing capacity and cannot perform complex image recognition tasks. They need to offload computational tasks to nearby devices that have the processing capacity to execute the tasks. These image processing tasks must be done in a time-critical manner since fire and pest detection are time sensitive tasks. 

In \cite{Zhao2020}, Zhao et al. proposed using a network that consists of multiple hovering UAVs and Multi-Access Edge Computing (MEC) devices to aid farm monitoring. The UAVs provided: connectivity between the IoT devices, and a central processing unit (CPU) that can perform the image classification task. The UAVs have mounted CPUs and can compute the image processing tasks themselves or they can forward the task to a nearby MEC device to compute the task. Although our architecture is similar, \cite{Zhao2020} focuses on optimal placement of UAVs. 

UAVs are battery operated devices and are therefore limited to a finite amount of energy and CPU capacity. The MEC device can assist the UAVs by taking over some of the computationally heavy tasks. MEC devices have a more powerful CPU, and can share the load of the UAVs so that the UAVs do not need to drain their battery quickly. Offloading some tasks to MEC will also prevent a bottleneck situation where all of the tasks are waiting to be executed in the UAV's processing queues. In this paper, we aim to extend the UAV's energy level because we want the UAVs to hover over the smart farm for as long as possible. We also aim for the tasks to be completed before their deadline.  

\par In this study, we introduce a problem that has two objectives: $i)$ extend the battery life of the UAVs, and $ii)$ minimize the number of delay violations. The longer the aerial base stations (ABS) can hover, the more they can assist the IoT devices in computing the intensive tasks. The UAVs also need to ensure that these time-sensitive tasks will be computed before their deadlines. 

The remainder of this paper is organized as follows. The related works are explained in Section II. Then, we detail the system model and the problem definition in Section III. Section IV introduces the proposed method and the baselines. Section V provides the computational experiment. Finally, in Section VI, we conclude the paper.

\section{Related Works}
\par There have been many studies emerging recently that suggest using UAVs in a smart farm. Zhou et al. provided an extensive survey for the use of UAVs with MEC in \cite{Zhou2020}. Lottes et al. detailed the usage of UAVs in a smart farm in \cite{Lottes2017}. They presented a real-world smart farm case in which a UAV device is used for image classification. In another study, Islam et al. introduced pesticide usage for a similar scenario and discuss the tradeoff between latency and battery usage \cite{Islam2021}. Lastly, Zhao et al. aimed to increase the throughput by considering the delay critical tasks in their smart farm environment \cite{Zhao2020}.
\par Energy efficiency and time critical task approaches for UAVs are not limited to smart farm scenarios. Yang et al. dealt with resource allocation and task offloading problems to reduce the overall power consumption in their networks \cite{Yang2019}. In another study, Zhou et al. included satellite computation/communication as an alternative for UAVs \cite{Zhou2021}. They modelled their problem as a constrained Markov decision process (MDP) and provided a deep reinforcement learning solution. Meanwhile, Ghdiri et al. provided a cluster-based approach for deadline aware task computation in UAVs \cite{Ghdiri2020}. Lastly, Yao et al. prefered a generalized Nash equilibrium approach for the offloading problem in a UAV swarm \cite{Yao2020}. Unlike the previous works, we focus on tasks deadlines and maximizing hover time by providing an integer linear programming (ILP) model and Q-Learning solution.

\begin{table}
\centering
\caption{\label{tab:Notations} Summary of the notations.}
\begin{tabular}{c|c|p{4cm}}
\textbf{Sets} & \textbf{Size} & \textbf{Description} \\ \hline
$t\in\mathcal{T}$ & $T$ & Set of time intervals \\
$j\in\mathcal{J}$ & $J$ &Set of UAVs  \\
$l\in\mathcal{L}$ & $L$ & Set of MEC devices \\
$j'\in\mathcal{J}^{+}$ & $J$+$L$ &Set of CPUs \\
$k\in\mathcal{K}$ & $K$ & Set of task types  \\ \hline
\textbf{Variables}& \textbf{Domain} & \textbf{Description} \\ \hline
$x_{jtj'}$ & $\{0,1\}$ & Task offload decision \\
$p_{jtj't'}$ & $\{0,1\}$ & CPU allocation \\
$p^{+}_{jtj't'}$ & $\{0,1\}$ & CPU allocation first time interval\\
$p^{-}_{jtj't'}$ & $\{0,1\}$ & CPU allocation last time interval \\
$v_{jt}$ & $\{0,1\}$ & Deadline violation  \\ \hline
\textbf{Task Parameters}& \textbf{Range} & \textbf{Description} \\ \hline
$\alpha^{B}_{jt}$ & $\{0,1\}$ & Task demand indicator \\
$\alpha^{P}_{jt}$ & $\mathbb{R}$ & Processing time \\
$\alpha^{D}_{jt}$ & $\mathbb{R}$ & Deadline \\ 
$\mathcal{V}^{L}_{j}$ & $\mathbb{N}$ & Violation reward level\\
$\Delta_{jt}$ & $\mathbb{R}$ & Scheduling + processing delay\\ \hline
\textbf{Energy Parameters}& \textbf{Range} & \textbf{Description} \\ \hline
$\Upsilon^{B}_{j}$ & $\mathbb{N}$ & Battery capacity \\
$\Upsilon^{H}_{j}$ & $\mathbb{R}$ & Hovering energy cons. \\	
$\Upsilon^{A}_{j}$ & $\mathbb{R}$ & Antenna energy cons. \\	
$\Upsilon^{I}_{j}$ & $\mathbb{R}$ & CPU idle energy cons. \\
$\Upsilon^{C}_{j}$ & $\mathbb{R}$ & CPU active energy cons. \\
$\Upsilon^{R}_{j}$ & $\mathbb{R}$ & Remaining energy in a battery\\
$\Upsilon^{L}_{j}$ & $\mathbb{N}^{+}$ & Battery reward level\\
$W$ & $[0,1]$ & Energy consumption weight\\
$\Theta$ & $\mathbb{R}$ & Scaling factor\\
\hline
\end{tabular}
\end{table}

\section{System Model}
\par We consider a set of UAVs, $j\in\mathcal{J}$, is hovering above a rural area, communicating with IoT devices and providing them with guaranteed service. In addition, a set of MEC servers  $l\in\mathcal{L}$,  is located at the edge of the smart farm and is available for task processing. One of the primary advantages of sharing the tasks with a MEC server is the extension of the hovering time of the UAVs, which have limited battery capacities ($\Upsilon^{B}_{j}$). Overall, CPUs ($j'\in \mathcal{J}^{+}$), located either in a UAV or MEC, can be selected to process the IoT tasks.

\par In our time interval based ($t\in \mathcal{T}$) model, the IoT devices may demand to process $K$ types of tasks from their associated UAV, $j\in \mathcal{J}$, in any of these time intervals ($\alpha^{B}_{jt}$). Each task type has a unique CPU processing time ($\alpha^{P}_{jt}$), and a deadline ($\alpha^{D}_{jt}$) that should not be violated to provide a reasonable quality of service to these IoTs. We propose a task offloading problem that will complete the tasks before their deadline, and improve the hovering time of the UAVs. The problem will be explained in the following subsection. 

\subsection{Task Offloading Optimization for Increasing UAV Hover Time and Reducing the Deadline Violations}
\par We combine our two significant key performance indicators (KPIs): increasing the hover time and reducing the task deadline violations, as a multi-objective maximization problem in Eq.~\ref{eq:obj}. Here, $W$ is the weight of the increasing hover time goal. $\Theta$ is a scale value used to normalize energy consumption (Watts), and the total number of deadline violations ($v_{jt}$). In order to improve overall hover time, we maximize the minimum remaining energy ($\Upsilon^{R}_{j'}$) in the batteries of the set of UAVs ($j'\in\mathcal{J}$). Therefore we can extend the operation time of all UAVs without the need to recharge their batteries. Eq.~\ref{eq:encalc} calculates the remaining energy in a battery by subtracting the hovering ($\Upsilon^{H}_{j'}$), antenna ($\Upsilon^{A}_{j'}$), and CPU idle ($\Upsilon^{I}_{j'}$) energy consumptions \footnote{Note that UAVs also need to consume energy for their other operations, such as communicating with other devices. However, the energy consumption associated with these are not impacted by our decisions in this system model. Thus, we do not include them to improve the readability of the problem definition.} from the full battery capacity ($\Upsilon^{B}_{j'}$), respectively. Lastly, we calculate the total CPU active energy consumption by multiplying the total number of time intervals that CPU $j'$ spent to process a task with the difference of CPU active ($\Upsilon^{C}_{j'}$) and idle energy consumptions.

\par Eq.~\ref{eq:cpualloc} ensures that each task generated in this smart farm ($\alpha^{B}_{jt}$) is allocated to one of the CPUs in this network. For that purpose, this equation includes two critical decision variables. First, the task offloading binary indicator $x_{jtj'}$, is equal to one if the CPU $j'$ processes the task received by UAV $j$ in time interval $t$. Second, the binary indicator $p_{jtj't'}$, is equal to one if that task is processed in the time interval $t'$. Note that if the required processing time ($\alpha^{P}_{jt}$) is longer than one time interval, multiple time intervals need to be allocated for that task. Lastly, the time intervals earlier than the task arrival time ($t'< t$) can not be allocated in a CPU to process this task (Eq.~\ref{eq:cpunegalloc}). A CPU's core can serve at most one task in a single time interval (Eq.~\ref{eq:cpuslotlimit}). However, it could be possible to extend this model to multitasking CPUs without loss of generality. Moreover, a task can be processed at most by one CPU in the same time interval (Eq~\ref{eq:uavlimit}). Lastly, we have to limit the number of offloading decisions to process the entire task in the same CPU (Eq~\ref{eq:offloadlimit}).

\begin{flalign}
&\textbf{Maximize:} \notag \\
\label{eq:obj}
&W * \min_{ j' \in J} \Upsilon^{R}_{j'} - \frac{1-W}{\Theta} \sum\limits_{\substack{j\in\mathcal{J}\\ t\in\mathcal{T}}}v_{jt}\\
\label{eq:encalc}
&\Upsilon^{R}_{j'} = \Upsilon^{B}_{j'} - (\Upsilon^{H}_{j'} + \Upsilon^{A}_{j'} + \Upsilon^{I}_{j'}) * T - \sum\limits_{\substack{j\in\mathcal{J}\\ t\in\mathcal{T} \\ t'\in\mathcal{T}}} (\Upsilon^{C}_{j'} - \Upsilon^{I}_{j'}) \notag \\&* p_{jtj't'} \\
&\textbf{Subject to:} \notag \\
    \label{eq:cpualloc}
	&\sum\limits_{t'=t}^{T}\sum\limits_{j' \in \mathcal{J}^{+}} p_{jtj't'} * x_{jtj'} = \alpha^{B}_{jt} * \alpha^{P}_{jt} ,\quad \forall j\in\mathcal{J} , \forall t\in\mathcal{T} \\
	\label{eq:cpunegalloc}
	&\sum\limits_{t'=0}^{t-1}\sum\limits_{j' \in \mathcal{J}^{+}} p_{jtj't'} * x_{jtj'} = 0 ,\qquad\qquad \forall j\in\mathcal{J} , \forall t\in\mathcal{T} \\
	\label{eq:cpuslotlimit}
	&\sum\limits_{j \in \mathcal{J}}\sum\limits_{t \in \mathcal{T}} p_{jtj't'}  \leq 1,\quad\qquad\qquad\qquad
	\sum\limits_{j \in \mathcal{J}}\sum\limits_{t \in \mathcal{T}} p^{+}_{jtj't'}  \leq 1, \notag\\
	&\sum\limits_{j \in \mathcal{J}}\sum\limits_{t \in \mathcal{T}} p^{-}_{jtj't'}  \leq 1 ,\quad\qquad\qquad\qquad \forall j'\in\mathcal{J}^{+} , \forall t'\in\mathcal{T} \\
	\label{eq:uavlimit}
	&\sum\limits_{j' \in \mathcal{J}^{+}} p_{jtj't'}  \leq 1 ,\quad\qquad\qquad\qquad\qquad 
	\sum\limits_{j' \in \mathcal{J}^{+}} p^{+}_{jtj't'}  \leq 1 ,\notag \\
	&\sum\limits_{j' \in \mathcal{J}^{+}} p^{-}_{jtj't'}  \leq 1 ,\quad\qquad\qquad \forall j\in\mathcal{J}, \forall t'\in\mathcal{T}, \forall t\in\mathcal{T} \\
	\label{eq:offloadlimit}
	&\sum\limits_{j'\in\mathcal{J}^{+}} x_{jtj'}  \leq 1 ,\quad\qquad\qquad\qquad\qquad \forall j\in\mathcal{J}, \forall t\in\mathcal{T} \\
	\label{eq:scheduling_delay}
	& \Delta_{jt} = \sum\limits_{\substack{j'\in\mathcal{J}^{+} \\ t'\in\mathcal{T}}} \left[p^{+}_{jtj't'} * (t') - t + \alpha^{P}_{jt} \right] \\
	\label{eq:delayviol1}
	&\alpha^{D}_{jt} \geq  \Delta_{jt} - M * v_{jt}, \quad\qquad\qquad\qquad\forall j\in\mathcal{J}, \forall t\in\mathcal{T} \\
	\label{eq:delayviol2}
	&\alpha^{D}_{jt} \leq \Delta_{jt} + M * (1-v_{jt}), \forall j\in\mathcal{J}, \qquad\qquad\forall t\in\mathcal{T}\\
	\label{eq:contiguity}
	&p_{jtj'(t'+1)} = p_{jtj't'}+p^{+}_{jtj'(t'+1)} - p^{-}_{jtj'(t'+1)} \\
	\label{eq:contiguity2}
	&p^{+}_{jtj'(t'+1)} + p^{-}_{jtj'(t'+1)} \leq 1 \notag\\
	&\forall j\in\mathcal{J}, \forall t\in\mathcal{T}, \forall j'\in\mathcal{J}^{+}, \forall t' \in \mathcal{T}  \\
    \label{eq:contiguity3}
	&p_{jtj'(0)} = p^{+}_{jtj'(0)}  \\
	\label{eq:contiguity4}
	&\sum\limits_{\substack{t'\in\mathcal{T}}} p^{+}_{jtj't'} \leq 1, \qquad\qquad \forall j\in\mathcal{J}, \forall t\in\mathcal{T}, \forall j'\in\mathcal{J}^{+} \\
	\label{eq:contiguity5}
	&\sum\limits_{\substack{t'\in\mathcal{T}}} p^{-}_{jtj't'} \leq 1, \qquad\qquad \forall j\in\mathcal{J}, \forall t\in\mathcal{T}, \forall j'\in\mathcal{J}^{+}
\end{flalign}

\par In order to calculate the summation of the scheduling and processing delay ($\Delta_{jt}$) of a task that arrived to UAV $j$ in time interval $t$ in Eq~\ref{eq:scheduling_delay}, we introduce a new decision variable, $p^{+}_{jtj't'}$, to render the first time interval, $t'$, that we processed this task in CPU $j'$. If this binary decision variable equals to one, CPU $j'$ starts to process that task. After we multiply that decision variable with $t'$ and subtract the task arrival time ($t$), we can then find the scheduling delay for this task. After adding the processing delay of this task ($\alpha^{P}_{jt}$) in to that equation, we can find the  summation of the scheduling and processing delay. 
\par We use the delay calculation to identify the deadline violation variable ($v_{jt}$). The association between that binary decision variable and the deadline of the task accomplished with Eqs.~\ref{eq:delayviol1} and \ref{eq:delayviol2} are done by using the Big-M method which is a common linear programming solving method that uses a very large constant in a constraint. In the case of a higher deadline value, $v_{jt}$ should be zero to satisfy the constraint defined by Eq.~\ref{eq:delayviol1} . Otherwise, $v_{jt}$ should be one to satisfy Eq.~\ref{eq:delayviol2}.         

\par As mentioned before, $p^{+}_{jtj't'}$ equals to one if the task arrives at the UAV $j$ in time interval $t$ and starts to be processed in $j'$ in time interval $t'$. Meanwhile, $p^{-}_{jtj't'}$ equals to one if the same task is completed in time interval $t'$. Therefore we can identify the exact start time and completion time of a task in a CPU. In addition, we want to process a task in a CPU without an interruption which is called contiguity.  Eqs.~\ref{eq:contiguity}-\ref{eq:contiguity5} ensure that CPU time interval allocation contiguity. If we detail these equations, Eq.~\ref{eq:contiguity} ensures that $p^{+}_{jtj't'}$ or $p^{-}_{jtj't'}$ should be one in a case when the values of consecutive allocation variables ($p{jtj't'}$ and $p{jtj'(t'+1)}$) are different, which actually means that we start to process the related task or finish to process it, respectively. Eq.~\ref{eq:contiguity2} eliminates the ping-pong effect. Eq.~\ref{eq:contiguity3} ensures that $p^{+}_{jtj't'}$ equals to one in case that we start to process the task in the first time interval. Lastly, Eq.~\ref{eq:contiguity4} and Eq.~\ref{eq:contiguity5} limit the start and completion time of a task, respectively. 

\section{Proposed Method}
\subsection{Q-Learning Approach}
\par We propose a finite-horizon multi-agent MDP framework to solve the problem explained in the previous section. Each UAV independently make decisions with a tuple $\mathcal{F} = \{\mathbb{P}, \mathbb{A}, \mathbb{R}, \mathbb{S}, \Pi \}$ in which:
\begin{itemize}
    \item \textbf{State Transitions:} $\mathbb{P}:\mathbb{S}_{1}x\mathbb{A}x\mathbb{S}_{2}\implies \mathbb{R}$
    \par The framework has a task-based state transition model. After each task arrives at the UAV, we update the environment and calculate the delays and battery levels to find the first state $\mathbb{S}_{1}$. After the action, we update the environment to find $\mathbb{S}_{2}$ and then reward $\mathbb{R}$. Due to unpredictable task arrivals, state transitions are stochastic. 
    \item \textbf{Action:} $\mathbb{A}= \{x_{j'\in\mathcal{J^{+}}}\}$
    \par After a UAV receives a task, it has three options: processing that task locally, offloading it to another UAV, or offloading the task to the MEC. That decision-making process is accomplished by choosing a CPU in the set of CPUs ($\mathcal{J}^{+}$) that includes all of these three alternatives.
    \item \textbf{State:} $\mathbb{S}= \{k, \Delta_{j'\in\mathcal{J^{+}}}, \Upsilon^{L}_{j'\in\mathcal{J}}$\}
    \par Here $k$ is the type of the task received by the UAV, $\Delta_{j'\in\mathcal{J^{+}}}$ are the delays in all CPUs, and $\Upsilon^{L}_{j'\in\mathcal{J}}$ is the battery levels of all UAVs calculated by Eq.~\ref{eq:rew2}. 
    \item \textbf{Policy($\Pi$):} We use an epsilon-greedy policy in this framework.
    \item \textbf{Reward:} 
    \begin{flalign}
    \label{eq:rew1}
    &\mathbb{R} = (\Upsilon^{L}_{j_{a}} - 1) + (1-\mathbb{E}(v_{j_{a}})) + \mathcal{V}^{L}_{j_{a}} * \mathbb{E}(v_{j_{a}})) \\
    \label{eq:rew2}
	& \Upsilon^{L}_{j_{a}} =
	\begin{cases} 
	        2, & \text{if } \mathbb{E}(\Upsilon^{R}_{j_{a}}) - \max_{j'\in\mathcal{J}}(\mathbb{E}(\Upsilon^{R}_{j'})) \ge -\epsilon \\
            0,& \text{if } \mathbb{E}(\Upsilon^{R}_{j_{a}}) - \max_{j'\in\mathcal{J}}(\mathbb{E}(\Upsilon^{R}_{j'})) \le -2*\epsilon \\
            1, & \text{otherwise,}
	\end{cases}\\
	\label{eq:rew3}
	& \mathcal{V}^{L}_{j_{a}} =
	\begin{cases} 
        -40, & \text{if } \mathbb{E}(v_{j_{m}})) = 0 \\
        -20,& \text{if } \mathbb{E}(v_{j_{r}})) = 0 \\
        -10,& \text{if } \exists j'\in(\mathcal{J}/(j_{r}\cup j_{a}))(\mathbb{E}(v_{j'})) = 0 \\
        -1, & \text{otherwise,}
	\end{cases}
	\end{flalign}
    We introduce the battery reward level concept ($\Upsilon^{L}_{j_{a}}$) to map the action ($j_{a}$) into the reward function (Eq.~\ref{eq:rew1}). A battery reward may have three levels [0,1,2]; thus, we return a negative (-1) reward for the battery level 0, and we return a positive (1) reward\footnote{We assume that MEC's energy is supplied by the main electricity grid; therefore, we always return a positive reward in the case of MEC selection} for the battery level 2. That battery level is calculated by Eq.~\ref{eq:rew2}, in which $\mathbb{E}(\Upsilon^{R}_{j'})$ is a UAV battery's expected remaining energy when that UAV starts to process the delegated task. If the difference between  maximum energy level and the energy level of the selected UAV is lower than a certain level ($\epsilon$), we promote this action with a positive reward. Therefore we can balance the remaining energy levels of the UAVs and increase their  hovering time. Lastly, we also introduce a hysteresis approach to that calculation by adding an extra battery level for the energy differences between [-$\epsilon$,-2$\epsilon$]. 
    \par In addition, the reward function (Eq.~\ref{eq:rew1}) includes the expected deadline violation $\mathbb{E}(v_{j_{a}})$, which equals zero if the task does not yield to a deadline violation at the delegated CPU (UAV or MEC $j_{a}$). In that case, the reward function returns 1. If a deadline violation has occurred, the reward function returns the violation reward level ($\mathcal{V}^{L}_{j_{a}}$), calculated by Eq.~\ref{eq:rew3}. In that equation, we focus on the expected delay violations if we delegated the task to a different CPU instead of the original action ($j_{a}$). There can be several possible scenarios. The first case is when we could have prevented  a delay violation if the task had been sent and processed in the MEC. We want to encourage the tasks to be processed in the MEC if it has an idle CPU, therefore we return a significant penalty in that condition. The second case occurs if we could have prevented a delay violation if the task was never offloaded from the received UAV in the first place. The third case represents the condition if we do not expect a violation from a different offloading decision. Otherwise, if a deadline violation would have occurred in any action, we return a small penalty. The numeric values selected as rewards or penalties are empirically set.
\end{itemize}
In the following subsections, we explain the baseline schemes.

\subsection{Baseline Approaches}
\par \subsubsection{Round Robin (RR)}
Our first baseline is the simple Round Robin method in which, when a UAV needs to offload a task, it selects the other UAVs and the MEC in a round robin fashion. 

\par \subsubsection{Highest Energy First (HEF)}
In this algorithm, the offloading decision is based on the UAV's remaining battery level. The UAVs in the network regularly update one another with their current remaining battery levels and they store this information in a table. When a UAV receives a task from an IoT device, it scans the table and finds the UAV with the highest remaining energy level. Once it has found the UAV with the highest energy level, it checks the difference between the highest remaining energy level and its energy level. If the difference is greater than the 1\% threshold, then the UAV will offload the task to the UAV with the higher energy level. Otherwise, the UAV will compute the task locally. Since the MEC device has unlimited power capacity, we limit the UAV's ability to offload a task to MEC to 20\% of the time.

\par \subsubsection{Lowest Queue Time and Highest Energy First (QHEF)}

We improve the HEF scheme by adding the queue time to the offloading decision. 
Queueing delay is defined as the sum of the processing delays of all the tasks that are currently in the UAV's CPU queue.  After a UAV receives a task from the IoT device, it finds the UAV with the lowest queuing delay. 
It checks to see if the difference between the current UAV's queueing delay and the lowest queuing delay is greater than the threshold of 0.5 seconds. If it is greater than the queueing delay threshold, then that UAV will be a contender for an offloading destination. Then it will try and find a UAV that has the lowest queuing delay but higher remaining battery level. 
The difference between the possible offloading destination and current UAV's battery levels must be above an energy threshold of 1\%. If such a UAV exist, then it will offload the task to that UAV, else, it will add the task to the list of tasks that the current UAV will process locally. Hence, this algorithm considers both the UAVs' battery levels and queueing delay in the decision-making process.

\par \subsubsection{ILP Solver} 
\par In addition to the heuristics we explained above, we use GUROBI Solver \cite{GurobiOptimization2021} to find the optimum solution for our multi-objective maximization problem (Eqs.~\ref{eq:obj}-\ref{eq:contiguity5}). Despite the NP-hard property of our problem, that method could be used for only small solution space problems. We detail the findings in the following section.

\section{Performance Evaluation}
\par \subsection{Simulation Platform}
To simulate the UAV network in the smart farm, we used Simu5G which is a 5G network simulator library developed over Omnet++ \cite{omnetpp}. Simu5G contains modules that model the different nodes found in a 5G and LTE network following the 3GPP standards \cite{Nardini2020}. 

\begin{table}
    \centering
    \caption{\label{tab:EngParam} Energy consumption parameters.}
    \begin{tabular}{c|c|c|c|c}
    $\Upsilon^{B}_{j'}$&$\Upsilon^{H}_{j'}$&$\Upsilon^{A}_{j'}$&$\Upsilon^{I}_{j'}$&$\Upsilon^{C}_{j'}$\\
    \hline
    570&211&17&4320&12960\\

    \end{tabular}
\end{table}  
\subsection{Energy Consumption Parameters}
We use the parameters\footnote{Due to limited simulation time, we set the energy consumption level of idle and busy CPU periods to be at the level they would be if they ran for ten hours, to highlight the performance of methods in terms of energy consumption.} in Table \ref{tab:EngParam} and Equation \ref{eq:encalc} to model the remaining UAV energy in our simulations. We assume that we are using a battery that is similar to the one found in \cite{HSE}. To calculate the power consumption of hovering, we used eq. 2 from \cite{Dorling2017}, we also used their assumptions such that that the UAV will have: 4 rotors, fluid density of $1.204 kg/m^3$, rotor disc area of $0.2m^2$, frame's mass $M=1.5$kg, and battery and payload $m = 3$ kg. The equation from \cite{Dorling2017} is given in eq. \ref{eqn:flyingpower}, where $M$ is the mass of UAV in kg, $m$ is the mass of the battery and other payloads in kg, $g$ is gravity in Newtons, $\rho$ is the fluid density in $kg/m^3$, $\varsigma$ is the rotor disc area in $m^2$, and $n$ is the number of rotors. 
\begin{equation}
    \label{eqn:flyingpower} 
    \Upsilon^{H}_{j'} = (M + m)^\frac{3}{2}\sqrt{\frac{g^3}{2\rho\varsigma n}}
\end{equation}

\begin{table}
    \centering
    \caption{\label{tab:SimParams} Simulation parameters.}
    \begin{tabular}{l|@{}c@{}c@{}c@{}c@{}}
     Parameter & Value \\ 
    \hline
    $J$ & 4  \\
    \hline
    $L$ & 1  \\
    \hline
    \multirow{3}{*}{$\mathcal{K}$}&Fire detection (FD)\\
    &Pesticide detection (PD)\\&Growth monitoring (GM)\\
    \hline
    Task Type & \begin{tabular}{p{1.2cm}|p{1.2cm}|p{1.2cm}|p{1.2cm}} $(1/\lambda)\tablefootnote{Mean interarrival rate.}$ & $\alpha^{D}_{jt}$ & $\alpha^{P}_{jt} (UAV)$ & $\alpha^{P}_{jt} (MEC)$\\ \end{tabular} \\
    \hline
    FD & \begin{tabular}{p{1.2cm}|p{1.2cm}|p{1.2cm}|p{1.2cm}} 0.25s & 0.3s & 0.1s & 0.05s  \end{tabular} \\
    PD & \begin{tabular}{p{1.2cm}|p{1.2cm}|p{1.2cm}|p{1.2cm}} 0.25s & 0.8s & 0.5s & 0.25s \end{tabular} \\
    GM & \begin{tabular}{p{1.2cm}|p{1.2cm}|p{1.2cm}|p{1.2cm}} 0.5s & 5s & 0.1s & 0.05s \end{tabular} \\
    \hline
    \end{tabular}
\end{table}

\par \subsection{Simulation Results} 
\begin{figure}
\centering
\includegraphics[width=0.45\textwidth]{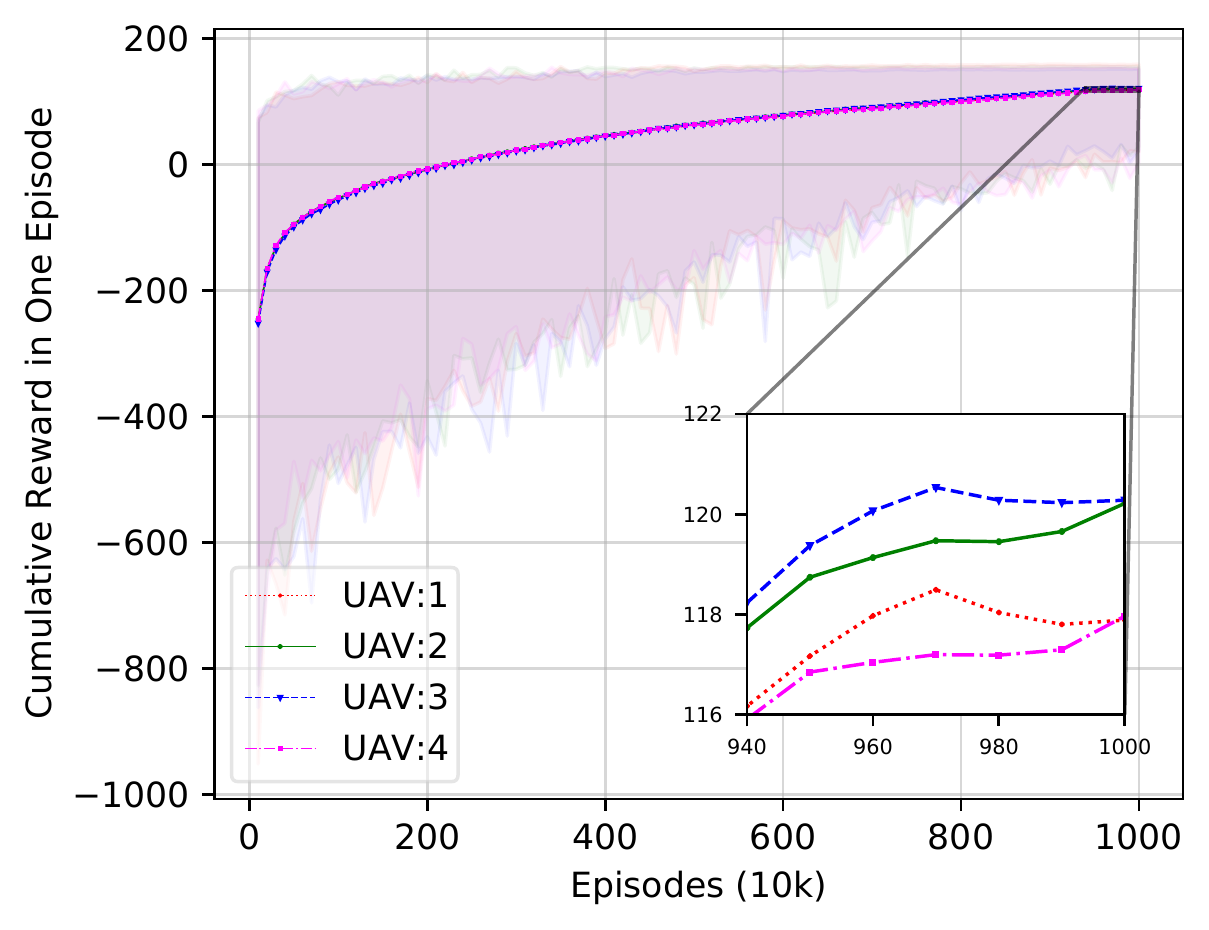}
\caption{\label{fig:obj} Convergence of multi-UAV Q-Learning method.}
\end{figure}

The results shown in this section are the average of ten runs with different seeds. Interarrival times between the tasks are exponentially distributed and their processing times are deterministic. Table \ref{tab:SimParams} presents the parameters used in all simulations. 
Fig.~\ref{fig:obj} shows the Q-Learning convergence in four UAVs. We used 0.05 as a learning rate, 0.85 for discount value and ran the scenario one million times (episodes). The solid lines show the average of 10k episodes, and the shaded area shows the variation of the cumulative reward. The average of the rewards in UAVs are close to each other in an episode, which means UAVs can learn independently with the same performance. Also, when we zoom in on the last 60k episodes, as seen in the subfigure, the difference in the cumulative reward is very low. 

\begin{figure}
    \centering
    \includegraphics[width=0.45\textwidth]{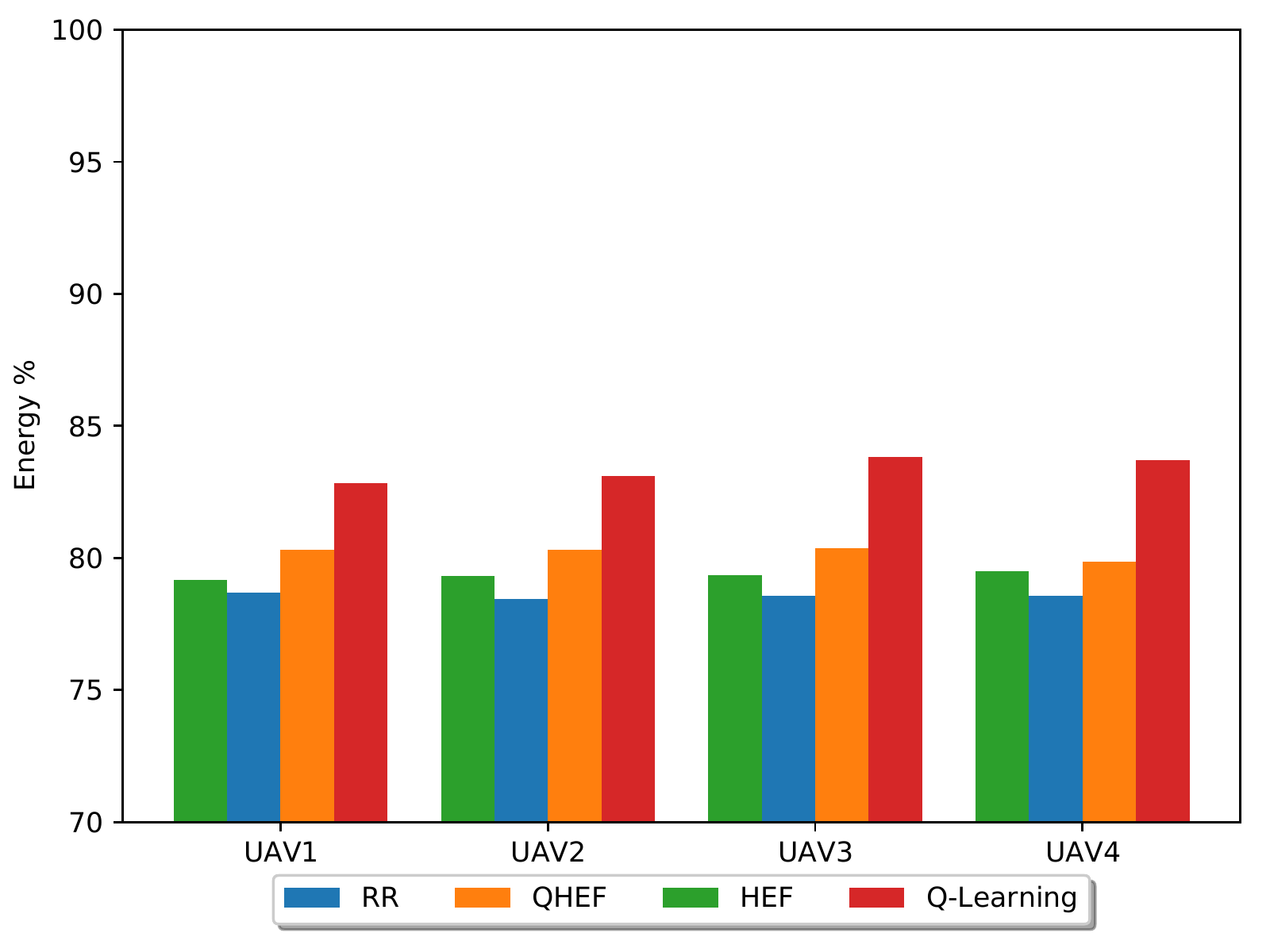}
    \caption{\label{fig:energy-plot} Comparison of remaining energy levels of UAVs.}
\end{figure}

The UAV's remaining energy percentage is the amount of energy left in the battery after the simulation has completed. The remaining energy percentage is directly related to the UAV's hover time. A higher remaining energy percentage means that the UAV will be able to hover over the network for a longer period of time. Fig.~\ref{fig:energy-plot} compares the remaining energy levels of the UAVs. The Q-Learning algorithm has the highest remaining energy level for all four of the UAVs.

In Q-Learning and QHEF, the UAVs offload their tasks to the MEC device more than the other UAVs. This allows the UAVs to not only preserve their own batteries, but also preserve the batteries of the other UAVs in the network. In HEF and RR, the majority of the offloaded tasks went to other neighbouring UAVs. Because the UAVs are not allowed to further offload an offloaded task, they must compute the offloaded task locally to prevent loops. Therefore, the additional offloaded tasks caused the UAVs to drain their batteries faster than the UAVs that used QHEF and Q-Learning.

\begin{figure}
    \centering
    \includegraphics[width=0.45\textwidth]{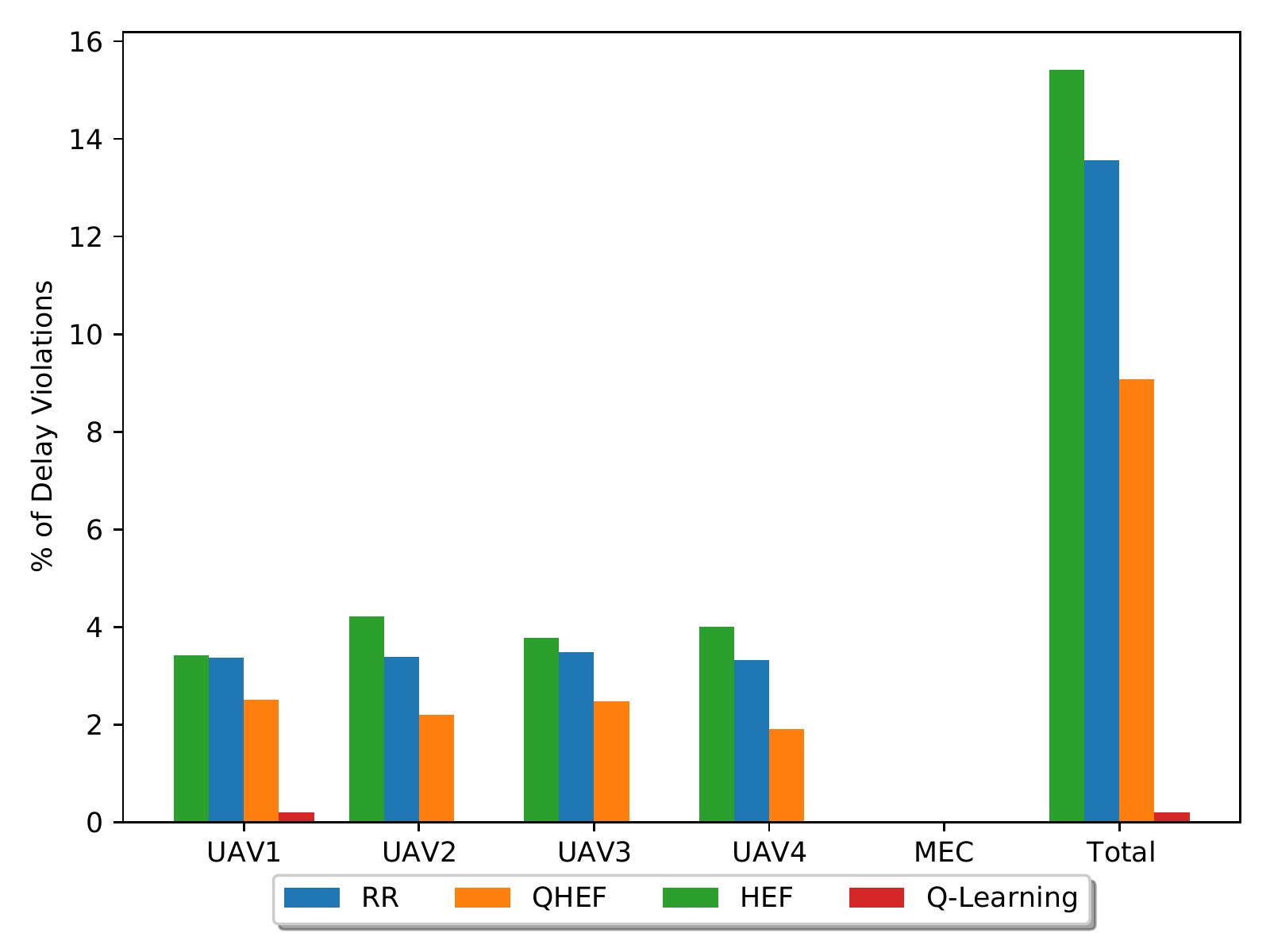}
    \caption{\label{fig:AllDV} Distribution of delay violations amongst smart farm nodes.}
\end{figure}

A delay violation occurs when the task has reached its predetermined deadline. Fig.~\ref{fig:AllDV} compares the percentage of delay violations out of the total number of tasks in the network. Q-Learning has the lowest percentages of delay violations, because it considers the task processing and scheduling delays in their decision-making process. QHEF selects the destination with the lowest queueing delay. Both of these algorithms consider the delay of a task, which is why they outperform RR and HEF in terms of delay violations.

\par In addition to the heuristic baselines, we used an ILP solver to find the upper bound for hover time and lower bound for delay violation KPIs. However, due to the problem's NP-hard property, we had to limit the simulation time to four seconds ($T=4s$) to find a solution with a 48 hours solver running time. In addition, we reduced the interarrival time to 0.125s for all task types and used more strict deadlines for fire detection ($0.2s$) and pesticide detection tasks ($0.6s$). 
\par Table~\ref{tab:ml} shows that the ILP solver can reduce the number of delay violations to seven if the solver only takes into account delay violations as the objective ($W=0$). On the other hand, when the solver focuses only on increasing the hover time ($W=1$), it can reduce the energy consumption in UAVs more than the other methods by offloading all tasks to MEC. However, that approach caused a significant rise in delay violations. It provided an adequate deadline violation if the solver is run as a multi-objective solver with a balanced value between these two KPIs ($W=0.5$). Lastly, if we compare the proposed Q-Learning method with this balanced solver method, Q-Learning provides a slight improvement in increasing the hover time, while it could not outperform the solver with regard to reducing the number of deadline violations. Nonetheless, the ILP solver needed 48 hours run-time which is significantly high.

\begin{table}
    \centering
    \caption{\label{tab:ml} Q-Learning and ILP solution comparison}
    \begin{tabular}{ c|c|c|c|c|c } 
	 Methods & $\Upsilon^{R}_{0}$& $\Upsilon^{R}_{1}$& $\Upsilon^{R}_{2}$&
	 $\Upsilon^{R}_{3}$&$\sum\limits_{\substack{j\in\mathcal{J}\\ t\in\mathcal{T}}}v_{jt}$\\
	 \hline
	 ILP ($W=0$) & $97.7\%$& $97.9\%$& $97.9\%$& $98.2\%$ &$7$\\
	 ILP ($W=1$) & $99.1\%$& $99.1\%$& $99.1\%$& $99.1\%$ &$24$\\
	 ILP ($W=0.5$) & $97.9\%$& $97.9\%$& $97.9\%$& $97.9\%$ &$9$\\
	 Q-Learning & $98.6\%$& $98.3\%$& $98.6\%$& $98.2\%$ &$13$\\ 
	 \hline
\end{tabular}
\end{table}

\section{Conclusion}
\par In this paper, we used a MEC assisted ABS network to provide a deadline aware service to the IoT tasks of a smart farm. In addition, we balanced the energy usage of the UAVs in this network to increase their hover time. We provided a Q-Learning based approach and several baselines to analyze our method. The results showed that our Q-Learning method performed better than the baselines in terms of remaining energy level and percentage of delay violations. Furthermore, ILP results demonstrated that the Q-Learning method is close to the optimum solution and exceeded the ILP approach by providing a solution for more extensive problems and being adaptive to changes in the environment. 

\section*{Acknowledgement}
This  work  is  supported  by  MITACS Canada Accelerate program under collaboration with Nokia Bell Labs.
\bibliography{tieraml}

\begin{thebibliography}{10}

\bibitem{Aldabbagh2020}
A.~D.~A. Aldabbagh, C.~Hairu, and M.~Hanafi, ``{Classification of Chili Plant
  Growth using Deep Learning},'' in {\em 10th International Conference on
  System Engineering and Technology (ICSET)}, IEEE, Nov 2020.

\bibitem{Lina2020}
Y.~Lina and Y.~Xiuming, ``{Design of Intelligent Pest Monitoring System Based
  on Image Classification Algorithm},'' in {\em 3rd International Conference on
  Control and Robots (ICCR)}, pp.~21--24, IEEE, Dec 2020.

\bibitem{Zhao2020}
J.~Zhao, Y.~Wang, Z.~Fei, and X.~Wang, ``{UAV Deployment Design for Maximizing
  Effective Data with Delay Constraint in a Smart Farm},'' in {\em
  International Conference on Communications in China (ICCC)}, IEEE, Aug 2020.

\bibitem{Zhou2020}
F.~Zhou, R.~Q. Hu, Z.~Li, and Y.~Wang, ``{Mobile Edge Computing in Unmanned
  Aerial Vehicle Networks},'' {\em IEEE Wireless Communications}, vol.~27,
  pp.~140--146, Feb 2020.

\bibitem{Lottes2017}
P.~Lottes, R.~Khanna, J.~Pfeifer, R.~Siegwart, and C.~Stachniss, ``{UAV-based
  crop and weed classification for smart farming},'' in {\em International
  Conference on Robotics and Automation (ICRA)}, IEEE, May 2017.

\bibitem{Islam2021}
N.~Islam, M.~M. Rashid, F.~Pasandideh, B.~Ray, S.~Moore, and R.~Kadel, ``{A
  Review of Applications and Communication Technologies for Internet of Things
  (IoT) and Unmanned Aerial Vehicle (UAV) Based Sustainable Smart Farming},''
  {\em Sustainability}, vol.~13, p.~1821, Feb 2021.

\bibitem{Yang2019}
Z.~Yang, C.~Pan, K.~Wang, and M.~Shikh-Bahaei, ``{Energy Efficient Resource
  Allocation in UAV-Enabled Mobile Edge Computing Networks},'' {\em IEEE
  Transactions on Wireless Communications}, vol.~18, pp.~4576--4589, Sep 2019.

\bibitem{Zhou2021}
C.~Zhou, W.~Wu, H.~He, P.~Yang, F.~Lyu, N.~Cheng, and X.~Shen, ``{Deep
  Reinforcement Learning for Delay-Oriented IoT Task Scheduling in SAGIN},''
  {\em IEEE Transactions on Wireless Communications}, vol.~20, pp.~911--925,
  Feb 2021.

\bibitem{Ghdiri2020}
O.~Ghdiri, W.~Jaafar, S.~Alfattani, J.~B. Abderrazak, and H.~Yanikomeroglu,
  ``{Energy-Efficient Multi-UAV Data Collection for IoT Networks with Time
  Deadlines},'' in {\em IEEE Global Communications Conference}, 2020.

\bibitem{Yao2020}
K.~Yao, J.~Chen, Y.~Zhang, L.~Cui, Y.~Yang, and Y.~Xu, ``{Joint Computation
  Offloading and Variable-width Channel Access Optimization in UAV Swarms},''
  in {\em IEEE Global Communications Conference}, 2020.

\bibitem{GurobiOptimization2021}
Gurobi, ``Optimizer reference manual,'' 2021.
\newblock Available online at:
  https://www.gurobi.com/documentation/9.1/refman/index.html, last accessed on
  2021-05-15.

\bibitem{omnetpp}
{OpenSim Ltd.}, ``{What is OMNeT++?},'' 2019.
\newblock Available online at: https://omnetpp.org/intro/, last accessed on
  2021-10-19.

\bibitem{Nardini2020}
G.~Nardini, D.~Sabella, G.~Stea, P.~Thakkar, and A.~Virdis, ``{Simu5G–An
  OMNeT++ Library for End-to-End Performance Evaluation of 5G Networks},'' {\em
  IEEE Access}, vol.~8, pp.~181176--181191, 2020.

\bibitem{HSE}
{HSE - UAV}, ``{High Power Drone Battery 6S-HV (LiHV)},'' 2021.
\newblock Available online at:
  https://hse-uav.com/product/high-power-drone-battery-6s-hv-lihv-25000mah-22-8v/,
  last accessed on 2021-10-19.

\bibitem{Dorling2017}
K.~Dorling, J.~Heinrichs, G.~G. Messier, and S.~Magierowski, ``{Vehicle Routing
  Problems for Drone Delivery},'' {\em IEEE Transactions on Systems, Man, and
  Cybernetics: Systems}, vol.~47, pp.~70--85, Jan 2017.

\end{thebibliography}

\end{document}